\documentclass[prl,twocolumn,superscriptaddress,floatfix,showpacs]{revtex4}

\usepackage{amsfonts,amsmath,amssymb,mathbbol}
\usepackage{latexsym}
\usepackage{graphicx}
\usepackage{epstopdf} 
\usepackage{bm,dcolumn}

\begin{document}

\title{Majorana Flat Bands in {\bf s}-Wave Gapless Topological Superconductors}

\author{Shusa Deng}
\affiliation{\mbox{Department of Physics and Astronomy, Dartmouth
College, 6127 Wilder Laboratory, Hanover, NH 03755, USA}}

\author{Gerardo Ortiz}
\affiliation{\mbox{Department of Physics, University of Indiana,
Bloomington, Indiana 47405, USA}}

\author{Amrit Poudel}
\affiliation{\mbox{Department of Physics and Astronomy, Dartmouth
College, 6127 Wilder Laboratory, Hanover, NH 03755, USA}}

\author{Lorenza Viola}
\affiliation{\mbox{Department of Physics and Astronomy, Dartmouth
College, 6127 Wilder Laboratory, Hanover, NH 03755, USA}}

\begin{abstract}
We demonstrate how the non-trivial interplay between spin-orbit coupling and 
nodeless $s$-wave superconductivity can drive a {\em fully gapped} two-band
topological insulator into a time-reversal invariant {\em gapless} topological 
superconductor supporting symmetry-protected Majorana flat bands.  We characterize 
topological phase diagrams by a ${\mathbb Z}_2 \times{\mathbb Z}_2$ partial Berry-phase 
invariant, and show that, despite the trivial crystal geometry, no {\em unique} bulk-boundary 
correspondence exists.  We trace this behavior to the {\em anisotropic} quasiparticle 
bulk gap closing, linear vs. quadratic, and argue that this provides a unifying principle 
for gapless topological superconductivity.  Experimental implications for 
tunneling conductance measurements are addressed, relevant for lead 
chalcogenide materials.  
\end{abstract}

\date{\today}
\maketitle

The emergence of ``topologically protected'' Majorana edge modes is a hallmark of 
topological superconductors (TSs) \cite{Book}.  Aside from their fundamental physical 
significance, Majorana modes are key building blocks in topological quantum 
computation \cite{Kitaev03}, due to their potential to realize non-Abelian braiding.  
As a result, a wealth of different approaches are being 
pursued theoretically and experimentally in the quest for topological quantum matter    
\cite{Book}, with recent highlights including broken time-reversal 
(TR) $p+ip$ superconductors, proximity-induced TR-invariant superconductivity in 
topological insulators (TIs), semiconductor-superconductor heterostructures, 
multiband superconductors and/or bilayer systems \cite{theory,Deng},
as well as experimental signatures of Majorana fermions in hybrid 
nanowires \cite{exp1} and doped TIs \cite{exp2}.  
Here, we propose a different paradigm, based on 
{\em topological gapless superconductivity in 
nodeless ($s$-wave) superconductors}. 

Gapless superconductivity is a physical phenomenon where the quasiparticle 
energy gap is suppressed (that is, it vanishes at particular momenta), 
while the superconducting order parameter remains {\em finite}, strictly non-zero.
This concept was anticipated on phenomenological grounds by 
Abrikosov and Gor'kov \cite{Abrikosov-Gorkov} in the context of 
TR pair-breaking effects in $s$-wave superconductors. 
Although certain unconventional superconductors may display similar behavior, their  
gapless nature results from the {\em nodal} character of the superconducting 
order parameter. In this work, the physical mechanism leading to a vanishing excitation 
gap is the spin-orbit coupling (SOC) in an otherwise {\em nodeless}, TR-invariant 
(centrosymmetric) multiband superconductor with bulk $s$-wave pairing. 

A consequence of such a state of matter is the emergence of 
surface {\em Majorana flat bands} (MFBs) if the spatial dimension $D\geq 2$.  
It has been appreciated that protected zero-energy flat bands may exist in 
unconventional nodal superconductors -- notably, at the surface of certain 
$d_{x^2-y^2}$-wave \cite{Hu}, $d_{xy}$-wave \cite{Sato2010}, and 
$d_{xy}+p$-wave superconductors \cite{Tanaka}; 
superconductors with a mixture of $d$- and $s$-wave pairing \cite{Schnyder}; 
$p\pm ip$ superconductors \cite{Wong} and superconducting helical magnets with 
effective $p$-wave pairing \cite{Martin} -- as well as in the vortex core 
of topological defects \cite{Volovik}.  Recently, a proposal for MFBs in 
nodeless $s$-wave (one-band) broken TR superconductors has also been put 
forward \cite{Vedral}. To the best of our knowledge, our model provides 
the first example of a TR-invariant $s$-wave gapless TS. We show that 
the {\em number} of Majorana edge modes in the non-trivial MFB phase 
(as opposed to just the {\em parity} of the number of Majorana pairs) is protected 
by a local chiral symmetry, a feature that is both crucial to understand 
robustness against perturbations and may be advantageous for 
topological quantum computation \cite{Alicea}.  
The dispersionless character of a MFB implies a large peak in the local density of 
states (LDOS) at the surface.  Thus, while detecting Majorana fermions through 
a zero-bias conductance peak in scanning tunneling microscopy (STM) 
experiments is not viable in gapped $D\geq 2$ TSs, an unambiguous experimental 
signature is predicted in the gapless case \cite{disorder,Vedral}.  

In addition to the above practical significance, an outstanding feature that our work 
unveils is the {\it anomalous, non-unique bulk-boundary correspondence} (BBC) that 
gapless TSs may exhibit: MFBs may emerge {\em only} along particular crystal directions, 
with {\em no} surface modes existing along others.  While such an anomalous BBC is 
reminiscent of the directional behavior typical of topological crystalline phases 
\cite{Fu11}, it does {\em not} stem simply from special crystal symmetries.  
Rather, the physical mechanism is rooted in the {\em anisotropic 
momentum dependence of the band degeneracy}: the quasiparticle gap may close 
non-linearly along certain directions, while it is linear (Dirac) along others. Only in the 
former case may a MFB exist at the corresponding edge. 
Accordingly, our findings suggest a {\em general} guiding principle for 
identifying and/or engineering materials supporting MFBs.  

\emph{Model Hamiltonian.---} We consider a two-band (say, orbitals $c$ and $d$) 
TR-invariant $s$-wave superconductor on a 2D square lattice.  
By letting  ${\bf k} \equiv (k_x,k_z)$ denote the wave-vector in the first Brillouin 
zone and $\psi_{\bf{k}}^\dag \equiv (c_{{\bf{k}},\uparrow}^\dag,
c_{{\bf{k}},\downarrow}^\dag,d_{{\bf{k}},\uparrow}^\dag,d_{{\bf{k}},
\downarrow}^\dag,c_{-{\bf{k}},\uparrow},c_{-{\bf{k}},\downarrow},d_{-{\bf{k}},
\uparrow},d_{-{\bf{k}},\downarrow})$, the relevant momentum-space Hamiltonian 
may be written as $H=\frac{1}{2}\sum_{\bf k}
\big( \psi_{\bf{k}}^{\dag} \hat{H}_{\bf{k}} \psi_{\bf{k}}-4\mu \big)$, 
where the $8\times 8$ matrix 
\begin{eqnarray}
\hat{H}_{\bf{k}}=s_z (m_{\bf k} \tau_z \hspace*{-0.5mm} - \hspace*{-0.5mm}\mu) + \tau_x 
({\lambda_{k_x}} \sigma_x \hspace*{-0.5mm}+ \hspace*{-0.5mm}{\lambda_{k_z}} \sigma_z)
-\Delta s_x \tau_y \sigma_x . \;
\label{Ham}
\end{eqnarray}
Here,  $s_\nu, \tau_\nu,\sigma_\nu$, $\nu= x,y,z$, are the Pauli matrices in the Nambu, 
orbital, and spin space, respectively, and tensor-product notation is understood. 
Physically, $m_{\bf{k}} \equiv u_{cd}-2t (\cos{k_x}+\cos{k_z})$, with 
$u_{cd}$ and $t$ representing the orbital-dependent on-site potential 
and the intraband hopping strength; $\mu$ is the chemical potential;  
${\lambda_{\bf k}} \equiv (\lambda_{k_x}, \lambda_{k_z})=
-2\lambda(\sin{k_x},\sin{k_z})$ describes the interband SOC, 
and $\Delta$ is the mean-field gap, with the superconducting 
pairing term being an interband $s$-wave spin-triplet of the form 
$H_{\text{sw}}= i\Delta \sum_j [ (c^\dag_{j,\uparrow} d^\dag_{j,\downarrow} +
c^\dag_{j,\downarrow} d^\dag_{j,\uparrow}) + \text{H.c.}]$, $\Delta \in 
{\mathbb R}$. 

\begin{figure}[t]
\includegraphics[width=8cm]{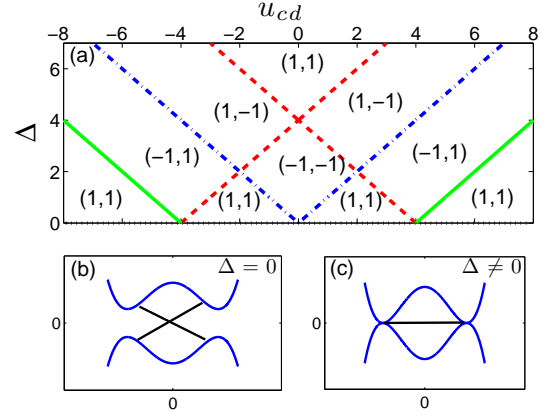}
\vspace*{-4mm} 
\caption{\label{pd} (Color online) Panel (a): Phase diagram of 
$H$ [Eq.~(\ref{Ham})] for $\mu=0= \lambda=1$. Each phase
is labelled by the partial Berry-phase parities $(P_{B,0}, P_{B,\pi})$. 
The topological numbers do not change under $\Delta \mapsto -\Delta$.
Panels (b) and (c): Sketch of the spectrum of $H$ with
$\Delta=0$ in a TI phase, and with $\Delta \ne 0$ in a TS flat-band
phase, respectively. }
\end{figure}

In addition to TR, particle-hole, and inversion symmetries \cite{Deng}, 
the Hamiltonian in Eq. (\ref{Ham}) obeys a special (unitary) chiral symmetry, 
$ [\hat{H}_{\bf{k}} , U_K ]_+=0$, where $U_K \equiv s_x \otimes \tau_z \otimes I$ 
and $I$ denotes the $2\times2$ identity matrix \cite{ChiralRemark}.
This symmetry will play an essential role in protecting MFBs.
We may decouple $\hat{H}_{\bf k}$ into two $4\times4$ blocks by 
applying a suitable unitary transformation $U$, followed by a 
reordering $P$ of the fermionic operator basis.  Specifically, 
let $U\hspace{-1mm}\equiv \hspace{-1mm}\frac{1}{\sqrt{2}}\{[
I \otimes (I+i\sigma_x)] \oplus \big[I \otimes (I-i\sigma_x)]\}$, 
with  
${P} \psi_{\bf k}^\dag \equiv (c_{{\bf{k}},\uparrow}^\dag,
d_{{\bf{k}},\downarrow}^\dag,
c_{-{\bf{k}},\downarrow},d_{-{\bf{k}},\uparrow},
c_{{\bf{k}},\downarrow}^\dag,d_{{\bf{k}},\uparrow}^\dag,
c_{-{\bf{k}},\uparrow},d_{-{\bf{k}},\downarrow})$.  Then 
$H$ is transformed into ${H'} =\frac{1}{2}\sum_{\bf{k}}
\,( \psi_{\bf{k}}^{\dag} \hat{H'}_{\bf{k}} \psi_{\bf{k}}^{\;}-4\mu )$, 
with $\hat{H'}_{\bf{k}} = (P U) \hat{H}_{\bf{k}} (P U)^\dag \equiv 
\hat{H'}_{1,\bf{k}} \oplus \hat{H'}_{2,\bf{k}}$. 
As in \cite{Deng}, $\hat{H'}_{1,\bf{k}}$ and $\hat{H'}_{2,\bf{k}}$ 
may be regarded as TR partners, and 
\begin{eqnarray*}
\hat{H'}_{1,\bf{k}}= \hspace{-1mm}\left (\hspace{-1mm}\begin{array}{cc}
m_{\bf{k}}\sigma_z-\mu+{\lambda_{\bf k}} \cdot \vec{\sigma}&
-i\Delta\sigma_y\hspace{-1mm} \\
\hspace{-1mm}i\Delta\sigma_y & -m_{\bf{k}}\sigma_z+\mu
+{\lambda_{\bf k}} \cdot \vec{\sigma}  \hspace{-1mm} \end{array} \right),
\label{4by4}
\end{eqnarray*}
with $\vec{\sigma} \equiv (\sigma_x,\sigma_y)$.
The exact quasiparticle excitation spectrum obtained by diagonalizing 
$\hat{H'}_{1,\bf{k}}$ is given by: 
\begin{eqnarray}
\epsilon_{n,{\bf
k}}\hspace{-0.5mm}=\hspace{-0.5mm}\pm\hspace{-0.5mm}
\sqrt{m_{\bf{k}}^2\hspace{-0.5mm}+
\hspace{-0.7mm}\Omega^2\hspace{-0.7mm}+\hspace{-0.5mm}|
\lambda_{\bf{k}}|^2\hspace{-0.7mm}\pm2\sqrt{
\mu^2\lambda_{k_x}^2\hspace{-0.5mm}+\hspace{-0.7mm}\Omega^2
(\lambda_{k_z}^2\hspace{-0.5mm}+\hspace{-0.5mm}m_{\bf{k}}^2)}},\;
\label{spectrum}
\end{eqnarray}
where we assume the order $\epsilon_{1,{\bf k}} \leq \epsilon_{2,{\bf k}} 
\leq 0 \leq \epsilon_{3,{\bf k}} \leq \epsilon_{4,{\bf k}}$, and $\Omega^2
\equiv \mu^2+\Delta^2$.  If no SOC is present, $\lambda=0$, then 
$\epsilon_{n,{\bf k}}=\pm(|m_{\bf{k}}| \pm |\Omega|)$, hence the gap
closes ($\epsilon_{2,{\bf k}}=0$) for $|m_{\bf{k}}|= |\Omega|$.  
By comparing $\epsilon_{2,{\bf k}}$ and  $\epsilon_{3,{\bf k}}$, one 
can see that as long as $|u_{cd}\pm 4t| > |\Omega| $, 
there is a {\em continuous} region of gapless bulk modes, which 
corresponds to a gapless two-band superconductor
with overlapping excitation spectrum \cite{Exp1}.  
If $\lambda \ne 0$, the situation is simplest 
at $\mu=0$, in which case 
$\epsilon_{n,{\bf k}}= \pm\, \sqrt{
\lambda_{k_x}^2 \hspace{-0.5mm}+\hspace{-0.5mm} 
(\sqrt{m_{\bf{k}}^2\hspace{-0.5mm}+\hspace{-0.7mm}
\lambda_{k_z}^2}\pm\Delta)^2 }$, 
and $\epsilon_{2,{\bf k}}=0$ when 
 $\lambda_{k_x}=0$, and $\lambda_{k_z}^2+m_{\bf
k}^2=\Delta^2$. For instance, if $\lambda=t\ne0$, 
this leads to $k_x \equiv k_{x,c} \in \{0,\pi\}$, and 
$k_z\equiv k_m =\pm \arccos \left(
\frac{(u_{cd}-2t\cos{k_{x,c}})^2+4t^2-\Delta^2}
{4 t (u_{cd} -2t\cos{k_{x,c}})} \right)$. 
Let $(k_x,k_z)\equiv (k_{x,c}, \pm k_m)$ denote the modes for which 
the bulk excitation spectrum closes.  We then expect only a 
{\em finite} set of values $k_m$ when $\lambda \ne 0$ for arbitrary 
$\mu$.  The quantum critical lines are 
determined by $\Delta=\pm m_{\bf k_c}$, with ${\bf k}_c \equiv 
(k_{x,c},k_{z,c})$ and $k_{z,c} \in \{0,\pi\}$) [Fig.~\ref{pd}(a)].

In the limit $\Delta=0$, our Hamiltonian reduces (up to unitary equivalence) to a 
TI model \cite{Franz}.  A qualitative comparison of the spectrum 
with open boundary conditions (OBC) along $\hat{z}$ with 
$\Delta=0$ vs. $\Delta \ne 0$ is shown in Fig.~\ref{pd}(b)-(c).  
Remarkably, we may consider our gapless TS to arise from doping  
a TI with {\em fully-gapped}, nodeless (spin-triplet) $s$-wave
superconductivity. More intuitively, an alternative route to realize our gapless TS is 
by turning on a suitable SOC in a two-band {\em gapless} superconductor, as the 
effect of $\lambda \ne 0$ is to separate the overlapping excitation spectrum 
and only leave a vanishing gap at a finite number of points. 
Thus, our nontrivial quasiparticle spectrum is a combined effect of SOC and 
superconducting order parameter.  The most striking aspect of such a 
spectrum is the fact that {\em the quasiparticle gap closing is anisotropic}: 
the gap vanishes {\em linearly} along $k_x$ [i.e., $\sim (k_x-k_{x,c})]$ and 
{\em quadratically} along $k_z$ [i.e., $\sim (k_z-k_{m})^2]$.  As we shall soon 
see, this peculiar behavior will manifest directly into an anomalous BBC. 

\emph{Topological response.---}
As a result of the gapless nature of the bulk excitation spectrum, 
topological invariants (such as the partial Chern number \cite{Deng}) 
applicable to 2D TR-invariant gapped TS systems are no longer 
appropriate. This motivates the use of {\em partial} 
Berry-phase indicators \cite{Deng}. 
In particular, we study the partial Berry phase of the two
occupied negative bands of one Kramers' sector only, $\hat{H}_{1,\bf{k}}$, 
for each $k_z$ (or $k_x$), namely, $B_{n,k_z}, n=1,2$, since the Berry phase 
of all the negative bands of $\hat{H}_{1,\bf{k}} $ and $\hat{H}_{2,\bf{k}}$ is
always trivial \cite{Deng}.  We can then compute the partial
Berry phase parity for each $k_z$ as
\begin{eqnarray}
P_{B,k_z}=(-1)^{{\rm mod}_{2\pi}(B_{+,k_z})/\pi}, \; B_{+,k_z} \equiv
B_{1,k_z}+B_{2,k_z},\;
\label{bnnp}
\end{eqnarray}  
and define a ${\mathbb Z}_2$ topological number as $\prod_{k_z}
P_{B,k_z}$. However, similar to the gapped case 
\cite{Deng}, the latter fails to identify 
quantum-critical lines between phases that share the same ${\mathbb Z}_2$ 
number.  For the purpose of identifying all the phase transitions 
and characterizing the {\em whole} phase diagram in Fig. \ref{pd}(a), a 
${\mathbb Z}_2 \times{\mathbb Z}_2$ indicator is necessary. 
Specifically, we define our topological invariant as $(P_{B,k_z=0},P_{B,k_z=\pi})$
[marked in each phase on Fig.~\ref{pd}(a)], which correctly signals a phase
transition whenever a jump of either $P_{B,k_z=0}$ or $P_{B,k_z=\pi}$ occurs. 
Since, as expected for a consistent bulk behavior, it turns out that  
$(P_{B,k_x=0},P_{B,k_x=\pi})=(P_{B,k_z=0},P_{B,k_z=\pi})$, 
we shall just write the ${\mathbb Z}_2 \times{\mathbb Z}_2$
invariant as $(P_{B,0},P_{B,\pi})$ henceforth. Note that while ultimately 
such a ${\mathbb Z}_2 \times{\mathbb Z}_2$
invariant involves only the partial Berry phase at ${\bf k}={\bf k_c}$, 
the reason for the more general definition of the topological numbers 
at $k_z \ne k_{z,c}$ is related to the BBC, as we discuss next.

\begin{figure}[tb]
\includegraphics[width=9cm]{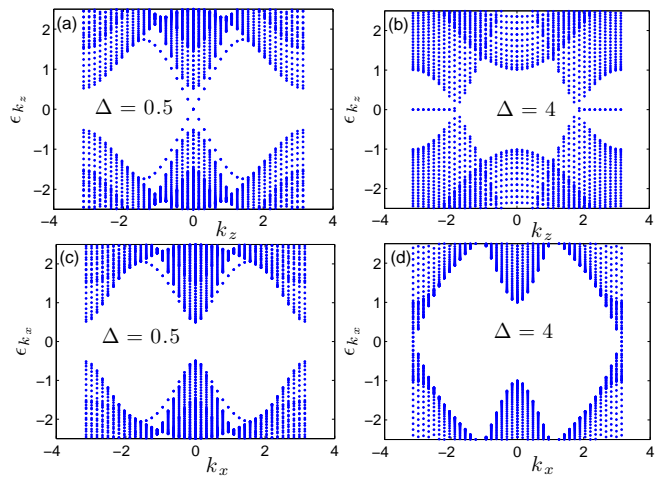}\vspace*{-3mm} 
\caption{\label{exud} (Color online) Excitation spectrum of $H$ [Eq. (\ref{Ham})] 
for $\mu=0, t=\lambda=u_{cd}=1$. Top (bottom) panels correspond to BC1 (BC2), 
whereas right vs. left columns correspond to 
$(P_{B,0}, P_{B,\pi})=(1,1)$ vs. $(1,-1)$. 
System size: $N_x=N_z=40$.}
\end{figure}

\emph{Bulk-boundary correspondence.---}
In a gapped TR-invariant TS, the BBC defines the relation between bulk 
topological invariants and the (parity of the) number of TR pairs of 
edge states \cite{Book,Ortiz11,Deng}. To understand the BBC in our gapless model, 
we contrast two situations:
BC1---periodic boundary conditions (PBC) along $\hat{z}$, and OBC
along $\hat{x}$; BC2--- PBC along $\hat{x}$, and OBC along $\hat{z}$.
Fig.~\ref{exud} shows how the excitation spectrum changes
as a function of $\Delta$ for BC1 (top panels) and BC2 (bottom panels)
for representative parameter choices in phases labelled by $(P_{B,0},
P_{B,\pi})=(1,1)$ [panels (a) and (c)], and $(P_{B,0},
P_{B,\pi})=(1,-1)$ [panels (b) and (d)]. In (a) there are two pairs of
Majorana modes on each boundary for $k_z=0$, but no Majorana edge modes 
in (c); likewise, in (b) there is a MFB for $k_m < |k_z| \leq \pi$ 
($k_m \approx 1.8$), but again no Majorana edge modes in (d). 
As further investigation under BC1 reveals, when $P_{B,k_z}=-1$ a {\em single}  
TR-pair of Majorana {\em edge} modes exists for that $k_z$-value on each
boundary. Thus, a MFB is generated when there is a dense set 
of $k_z$ for which $P_{B,k_z}=-1$. On the contrary, the partial
Berry phase for $k_x \neq k_{x,c}$ is always trivial ({i.e.},
$P_{B,k_x}=1$); and when $P_{B,k_{x,c}}=-1$, it corresponds to 
gapless {\em bulk} modes for that $k_{x,c}$. 

The above results demonstrate the asymmetry between
the $\hat{x}$ and $\hat{z}$ directions notwithstanding their geometrical 
equivalence -- in direct correspondence with the anisotropic momentum dependence 
of the bulk excitation gap, as anticipated \cite{Remark}. 
We stress that although the choice of Hamiltonian in Eq. (\ref{Ham}) is 
motivated by our earlier work \cite{Deng}, different physical realizations of $s$-wave 
gapless TR-invariant TSs  may be envisioned as long as a similar mechanism is in place: 
notably, we may change $H_{\text{sw}}$ to {\em interband $s$-wave spin-singlet}, 
$H'_{\text{sw}}= \sum_j \Delta [(c^\dag_{j,\uparrow} d^\dag_{j,\downarrow} - 
c^\dag_{j,\downarrow} d^\dag_{j,\uparrow}) + \text{H.c.}]$, while also ensuring 
that the strength of the SOC is sufficiently anisotropic, 
e.g., $(\lambda_{k_x}, \lambda_{k_z})=-2 (\lambda_x \sin{k_x},\lambda_z \sin{k_z})$, 
with $\lambda_z\ll \lambda_x$.  Based on these observations, we conjecture that the 
momentum asymmetry of the (bulk) excitation gap closing is a 
necessary condition for anomalous BBC, and that MFBs are necessarily associated 
with higher-than-linear closing.   Direct calculation confirms that this conjecture holds 
across a variety of models supporting surface flat bands: in particular, anomalous 
BBC is observed in spin-triplet $p_x+ip_y$ TSs \cite{Wong}, in both 
$s$-wave and $d_{x^2-y^2}$-wave spin-singlet TSs \cite{Vedral}, 
as well as a TR-broken TI model \cite{Dahm}. Interestingly, 
MFBs emerge along {\em both} spatial directions in $d_{xy}$ TSs 
\cite{Sato2010}, consistent with the symmetric (quadratic) closing of the  
bulk gap.

\begin{figure}[t]
\includegraphics[width=8.8cm]{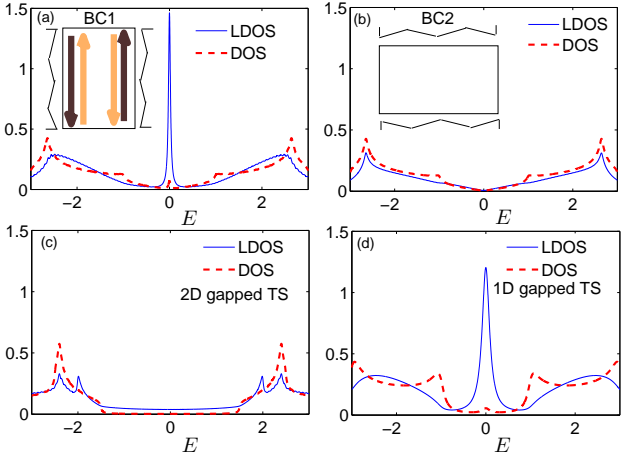}\vspace*{-3mm} 
\caption{\label{LDOS} (Color online) 
Panels (a) and (b): LDOS and DOS for $H$ [Eq. (\ref{Ham})] for 
$\mu=0, t=\lambda=u_{cd}=1$, $\Delta=4$. 
Insets: The jagged lines signify cuts of the system, implying 
that in BC1 (BC2) the OBC is along the $\hat{x}$ ($\hat{z}$). 
In (a), the brown (orange) arrows indicate a 
continuum of TR-pairs of Majoranas on each boundary, propagating along opposite 
directions. Panels (c) and (d): LDOS and DOS for a gapped TS in 2D and 1D. 
System size: $(N_x,N_z)=(80,400)$ (a), $(N_x,N_z)=(400,80)$ (b), 
$(N_x,N_z)=(80,400)$ (c), $N_x=80$ (d). }
\end{figure}

\emph{Observable signatures of Majorana flat band.---}
The tunneling current between a STM and the 
material is proportional to the surface LDOS of electrons \cite{Sarma12}. 
Results of LDOS calculations are shown in Fig.~\ref{LDOS}, 
together with the corresponding bulk density of states (DOS):
a huge (small) peak for the LDOS (DOS) is seen at zero energy under 
BC1 in (a), whereas no zero-energy peak occurs under BC2 in (b). While the quantitative 
difference between the LDOS vs. DOS peaks in panel (a) does indicates that the 
zero-energy modes are located on the boundary, the qualitative difference 
between panels (a) and (b) reinforces the asymmetric behavior under the two 
boundary conditions shown in Fig.~\ref{exud}.  It is instructive to 
compare to a typical gapped TS, e.g., the TR-invariant model discussed in 
Ref.~\cite{Deng}.  Although in this case Majorana edge modes exist 
in a nontrivial phase regardless of the direction along which OBC are assigned, 
{\em no} peak in LDOS (DOS) is seen at zero energy for $D>1$ [panels (c)-(d)]: 
in 2D (and 3D), the contribution to the LDOS from the finite 
number of Majorana edge modes is washed out by the extensive one from 
the bulk modes as the system size grows. 
Thus, a mechanism other than the existence of a finite number of Majoranas 
is needed to explain a 
zero-bias peak in $2$D ($3$D) fully-gapped superconductors. 

\emph{Robustness of Majorana flat band.---}  
Let us first consider a TR-preserving perturbation of the form 
$H_p =\sum_{j_x,k_z,k'_z,\sigma} u_p\,
(c_{j_x,k_z,\sigma}^\dag c_{j_x,k'_z,\sigma} + 
d_{j_x,k_z,\sigma}^\dag d_{j_x,k'_z,\sigma}) + \text{H.c.},$ 
where $k'_z \in \{ -k_z, \pi-k_z\}$, $u_p\in {\mathbb R}$.
Since $H_p$ allows Majorana modes at $k_z$ and $k'_z$ to 
couple with each other, it could significantly change the number of edge 
modes in principle. However, the zero-energy modes on
the left (right) boundary of 
$\hat{H}'_{1,k_z}$, 
say $\gamma_{k_z,\ell}$
($\ell=L,R$), may be taken to be eigenstates of ${\cal K}$, 
i.e., $\mathcal{K} \gamma_{k_z,\ell}=\pm  \gamma_{k_z,\ell}$, when 
there is only one edge mode on each boundary for $k_z$.
Thus, when there is {\em only one pair of zero-energy modes in the bulk}, 
at $k_z=\pm k_m$, all the zero-energy edge modes on the same
boundary can be continuously deformed one into another, 
which guarantees that they belong to the same sector of $\mathcal{K}$. Therefore, 
any local perturbation that preserves both chirality and TR  cannot 
lift the degeneracy of the zero-energy modes belonging to the
same sector of $\mathcal{K}$, leaving the MFB stable. 
However, {\em the protection from ${\mathcal K}$ 
may fail when there is an even number of pairs of zero-energy bulk modes}: 
e.g., in the phase $(P_{B,0},P_{B,\pi})=(-1,-1)$, the MFB is 
not robust against $H_p$, since now Majoranas on the same boundary 
may belong to different sectors of $\mathcal{K}$. 
Thus, not only does the {\em parity} of the number of
Kramers' pairs of Majoranas still play an important role, but also 
the {\em number} of edge modes in the MFB 
is conserved as long as both symmetries are respected and 
there is only one pair of bulk gapless modes.
Similarly, 
the MFB is robust
against another natural TR-preserving perturbation, namely, {\em intraband}
$s$-wave pairing, $H_s=\Delta_c \sum_ j (
c^\dag_{j,\uparrow}c^\dag_{j,\downarrow}+\Delta_d \sum_ j
d^\dag_{j,\uparrow}d^\dag_{j,\downarrow})+\text{H.c.}$, $\Delta_c, 
\Delta_d \in {\mathbb R}$, which anti-commutes with $U_K$. 

\begin{figure}[t]
\includegraphics[width=8.8cm]{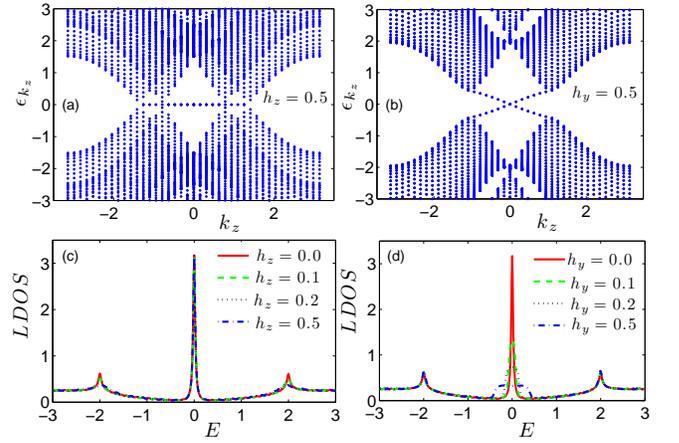}
\vspace*{-6mm}
\caption{\label{ldoshyzn} (Color online) Excitation spectrum of $H+H_\nu$ 
for $\mu=0, t=\lambda=1, u_{cd}=4, \Delta=2$. Panels (a) and (b): in-plane vs. 
out-of-plane field. Panels (c) and (d): 
LDOS for increasing field strength along $\hat{z}$ vs.
$\hat{y}$.  System size: $(N_x,N_z)=(40,40)$ [(a), (b)],
$(N_x,N_z)=(40,400)$ [(c), (d)]. }
\end{figure}

Next, consider TR-breaking perturbations due to a static magnetic field 
\cite{Deng,Eugene},
 $H_\nu=h_\nu \sum_j \psi_j^\dag \sigma_\nu \psi_j$,  
where $\nu=\hat{x},\hat{z}$ $(\hat{y})$ correspond to in-plane (out-of-plane) 
directions. 
The response to an in-plane field is similar in both 
directions, with the MFB remaining flat, Fig. ~\ref{ldoshyzn}(a). 
Under a magnetic field $h_y$, instead, the MFB becomes unstable, 
Fig.~\ref{ldoshyzn}(b). 
The effect of the magnetic field along different directions 
may be understood  through its relation with chirality. Specifically, the $\hat{x}$ 
and $\hat{z}$-components of $H_\nu$ anti-commute with ${\cal K}$, 
whereas $H_y$ {\em commutes} with ${\cal K}$.  Accordingly, chirality-protection is lost in this
case. The 
LDOS in the presence of Zeeman fields 
along in-plane ($\hat{z}$) and out-of-plane ($\hat{y}$) directions 
is shown in Fig.~\ref{ldoshyzn} (c)-(d): the peak at zero energy stays 
almost unchanged as $h_z$ increases, 
whereas it is strongly suppressed when $h_y \ne 0$. This is
consistent with the results from the excitation spectrum shown above.
Moreover, the behavior of the LDOS 
under a magnetic
field along an {\em arbitrary} direction on the $\hat{x}$-$\hat{z}$ plane is similar
to the one under $h_z$.  
We may then infer that a MFB responds to a uniform Zeeman 
field along a certain direction in a similar way as to a magnetic
impurity field along the same direction. Thus, the MFB 
will be robust in the presence of in-plane magnetic impurities, 
which may be unavoidable in real materials. 
Lastly, we investigated the effect of on-site disorder 
along the boundary,  $H_d=\sum_j v_j (c^\dag_{j, \uparrow} \, c_{j, \uparrow} + 
c^\dag_{j, \downarrow} \, c_{j, \downarrow} + $ $ d^\dag_{j, \uparrow} \, d_{j, \uparrow} + 
d^\dag_{j, \downarrow} \, d_{j, \downarrow} ) + \mbox{H.c}$, 
where $v_j \in {\mathbb R}$ is a Gaussian random potential.  The MFBs is 
robust against weak disorder so long as chirality is preserved, with 
the zero-energy peak in the LDOS remaining qualitatively intact. 

\emph{Conclusion.---} 
Majorana modes in gapless TSs can manifest themselves through new signatures, 
such as the emergence of a chirality-protected MFB which may depend crucially on the 
nature of the boundary.  
Such an anomalous, non-unique, BBC in $2$D ($3$D) gapless TSs allows for a more 
unambiguous signature in tunneling experiments than gapped TSs
may afford.  The anisotropic, linear vs. non-linear, vanishing of the 
quasiparticle bulk excitation gap at particular momenta 
is the  unifying principle behind such anomaly. Our model provides an 
explicit realization of a TR-invariant two-band gapless TS, where 
an anisotropic excitation spectrum arises from the interplay of conventional $s$-wave 
superconductivity with a SOC whose form is 
motivated by band-structure 
studies in Pb$_x$Sn$_{1-x}$Te \cite{Dimmock}. Thus, we expect that materials in this 
class may be natural candidates for the experimental search of TR-invariant gapped 
\cite{Deng} or gapless TSs. 

We thank Jake Taylor and Yuji Matsuda for 
discussions. Support from the NSF through grants No. PHY-0903727 
and PHY-1104403 (to LV) is acknowledged.


\begin{thebibliography}{100}
\bibitem{Remark} Interestingly, 
$\prod_{k_z} P_{B,k_z}=\prod_{k_x} P_{B,k_x}$, which indicates
consistent bulk properties as a whole, although the topological numbers
for each $k_x$, $k_z$ may differ individually.
\end{thebibliography}
\end{document}